\documentclass[twocolumn,aps]{revtex4}
 
\usepackage{epsf}
 
\begin{document}

\title{High-Order Coupled Cluster Calculations Via Parallel Processing: \\
 An Illustration For CaV$_4$O$_9$}

\author{D.J.J. Farnell$^1$, J. Schulenburg$^2$, J. Richter$^2$, and 
K.A. Gernoth$^3$, }

\affiliation{$^1$Unit of Ophthalmology, Department of Medicine, 
 University of  Liverpool, Liverpool, United Kingdom}

\affiliation{$^2$University of Magdeburg, Magdeburg, Germany}

\affiliation{$^3$School of Physics and Astronomy, University of Manchester, 
 Manchester, United Kingdom}

\date{\today}

\begin{abstract}
The coupled cluster method (CCM) is a method of quantum 
many-body theory that may provide accurate results for
the ground-state properties of lattice quantum spin systems
even in the presence of strong frustration and for lattices
of arbitrary spatial dimensionality. Here we present a 
significant extension of the method by introducing a new
approach that allows an efficient parallelization of 
computer codes that carry out ``high-order'' CCM calculations. 
We find that we are able to extend such CCM calculations 
by an order of magnitude higher than ever before utilized in a 
high-order CCM calculation for an antiferromagnet. Furthermore, we 
use only a relatively modest number of processors, namely, eight. 
Such very high-order CCM calculations are possible {\it only}
by using such a parallelized approach. An illustration 
of the new approach is presented for the ground-state properties 
of a highly frustrated two-dimensional magnetic material, 
CaV$_4$O$_9$. Our best results for the ground-state energy 
and sublattice magnetization for the pure nearest-neighbor
model are given by $E_g/N=-0.5534$ and $M=0.19$, respectively, 
and we predict that there is no N\'eel ordering in the region 
$0.2 \le J_2/J_1 \le 0.7$. These results are shown to be 
in excellent agreement with the best results of other 
approximate methods.  
\end{abstract}

\maketitle

\begin{figure}
\epsfxsize=6cm
\centerline{\epsffile{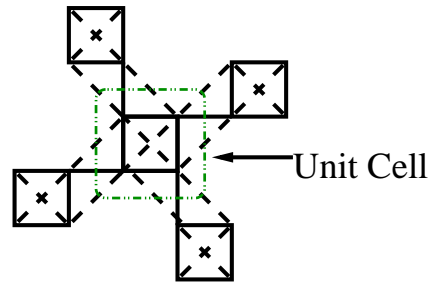}}
\vspace{-0.35cm}
\caption{The unit cell of this system contains four sites
and is indicated in the figure. Nearest-neighbor bonds are 
shown by the full lines and next-nearest-neighbor bonds by 
the dashed lines.}
\label{fig1}
\end{figure}

A new procedure for carrying out high-order coupled cluster 
method (CCM) \cite{ccm_theory1,ccm1,ccm2,ccm3,ccm4,ccm5,ccm6,xian} calculations 
via parallel processing is presented here. The CCM may be applied
to systems demonstrating strong frustration and for arbitrary 
spatial dimension of the lattice. We illustrate our 
new approach by applying it to CaV$_4$O$_9$ 
\cite{starykh,troyer,albrecht,weihong1,weihong2,ueda,miyazaki} 
at zero temperature. An increasing 
number of insulating quantum magnetic systems for lattices 
of low spatial dimensionality are being studied experimentally. 
Indeed, the calcium vanadium oxide (CAVO) materials 
are one particularly useful example. They exhibit 
strong frustration for a number of different crystallographic 
lattices with respect to varying chemical composition, 
and they may demonstrate ``novel'' ground-state ordering. 
Indeed, theoretical evidence suggests that CAVO systems 
may contain a number of differing quantum ground states (see, e.g., 
Refs. \cite{starykh,albrecht}) at zero temperature as a 
function of varying nearest-neighbor next-nearest-neighbor 
bond strengths. The relevant Hamiltonian is given by  
%%%%%%%%%%%
\begin{equation}
H = J_1 \sum_{\langle i,j \rangle } {\bf s}_i .
{\bf s}_j ~ + J_2 \sum_{\langle i,k \rangle } 
{\bf s}_i . {\bf s}_k ~ , 
\label{eq1}
\end{equation}
%%%%%%%%%%%
where $i$ runs over all lattice sites, and $j$ 
and $k$ run over all nearest-neighbor and  
next-nearest-neighbor sites to $i$, respectively, 
counting each bond once and once only. The bond 
strengths are given by $J_1$ and $J_2$ for nearest-neighbor 
and next-nearest-neighbor terms, respectively. The lattice and 
exchange ``bonds'' are illustrated graphically in Fig. \ref{fig1}.

For this model, collinear N\'eel ordering is observed 
with respect to nearest neighbors at $J_1>0$ and 
$J_2=0$ and with respect to next-nearest neighbors 
at $J_2>0$ and $J_1=0$. Exact diagonalizations 
\cite{albrecht}  predict an intermediate phase (or 
phases) for $0.2 < J_2/J_1 < 0.7$, which is also in 
good agreement with results of a  ``spin-wave theory-like'' 
treatment \cite{starykh} that predict such an intermediate regime 
for $0.25 < J_2/J_1 < 0.8$. Strong evidence has been gathered 
using approximate approaches \cite{starykh,weihong1,weihong2,ueda} 
that the ground state in this "intermediate" regime demonstrate dimer 
or plaquette ordering. However, conclusive evidence 
for the nature of the ground state across the whole 
of this regime remains to be yet determined. The ground 
state of the pure nearest-neighbor model at $J=1$ and 
$J_2=0$ demonstrates semi-classical N\'eel order
with a sublattice magnetization of $0.178(8)$ 
via QMC \cite{troyer}.

\begin{figure}
\epsfxsize=7cm
\epsffile{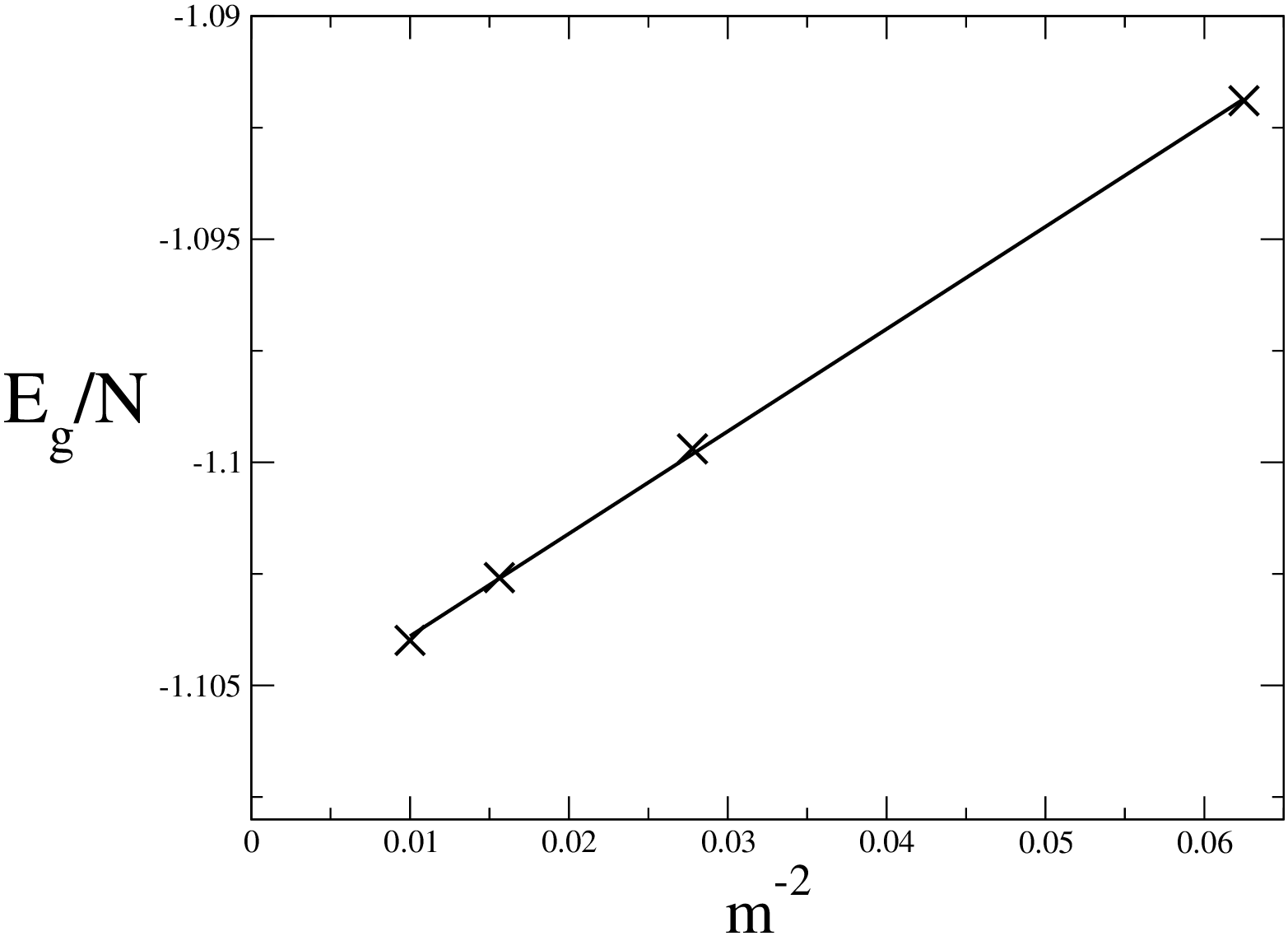}
\vspace{-3.75cm}
\caption{CCM results for the ground-state energy per spin 
at $J_1=1$ and $J_2=0$ using the LSUB$m$ approximation 
with $m=\{4,6,8,10\}$. The straight line is the linear
``best fit'' to the data.}
\label{fig2}
\end{figure}

We now turn our attention to the application of the CCM to 
this model. We choose two model states that are the classical
ground states of this model in the limits $J_2/J_1 \rightarrow 0$
and $J_2/J_1 \rightarrow \infty$ (with both $J_2$ 
and $J_1$ positive). These are, namely, one in 
which nearest-neighbor spins are antiparallel, henceforth 
denoted  the  nearest-neighbor N\'eel model state;  and a 
second model state in which next-nearest-neighbor spins are 
antiparallel, henceforth denoted the  next-nearest-neighbor 
N\'eel model state. We use four spins in the 
crystallographic primitive cell in both cases. This unit 
cell is indicated in Fig. \ref{fig1}. We perform a notational rotation of the 
``up'' spins to ``down'' spins, as in Refs. \cite{ccm1,ccm2,ccm3,ccm4,ccm5,ccm6,xian} 
for example. We use the full symmetries of the lattice 
to further reduce the computational problem of finding
 the fundamental CCM configurations and in determining 
and solving the CCM equations in both cases. 

\begin{figure}
\epsfxsize=7cm
\epsffile{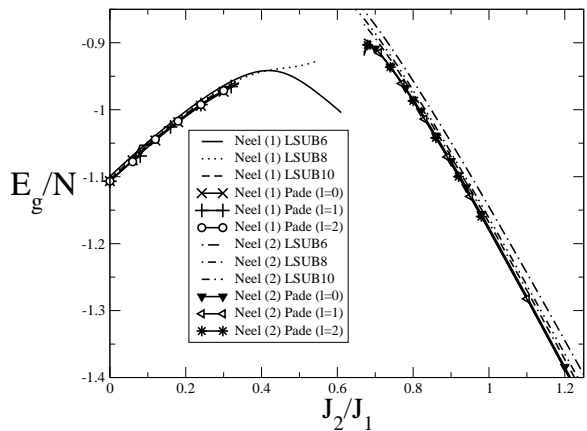}
\vspace{-3.75cm}
\caption{CCM results for the ground-state energy per spin of the 
$J_1$--$J_2$ model (with $J_1=1$) using the LSUB$m$ approximation 
with $m=\{6,8,10\}$. We see that our results converge 
rapidly with $m$ across a wide range of $J_2$ up to and including
the point $J_2=0$. Extrapolations using Pad\'e approximants 
are also shown.}
\label{fig3}
\end{figure}
 
The exact ket and bra ground-state energy eigenvectors, 
$|\Psi\rangle$ and $\langle\tilde{\Psi}|$, of a general 
many-body system described by a Hamiltonian $H$, 
%%%%%%%%%%%%%%%%%%%%%%%%%%%%%%%%%%
\begin{equation} 
H |\Psi\rangle = E_g |\Psi\rangle
\;; 
\;\;\;  
\langle\tilde{\Psi}| H = E_g \langle\tilde{\Psi}| 
\;, 
\label{ccm_eq1} 
\end{equation} 
%%%%%%%%%%%%%%%%%%%%%%%%%%%%%%%%%%  
are parameterized within the single-reference CCM as follows:   
%%%%%%%%%%%%%%%%%% 
\begin{eqnarray} 
|\Psi\rangle = {\rm e}^S |\Phi\rangle \; &;&  
\;\;\; S=\sum_{I \neq 0} {\cal S}_I C_I^{+}  \nonumber \; , \\ 
\langle\tilde{\Psi}| = \langle\Phi| \tilde{S} {\rm e}^{-S} \; &;& 
\;\;\; \tilde{S} =1 + \sum_{I \neq 0} \tilde{{\cal S}}_I C_I^{-} \; .  
\label{ccm_eq2} 
\end{eqnarray} 
%%%%%%%%%%%%%%%%%% 
The single model or reference state $|\Phi\rangle$ 
is required to have the property of being a cyclic vector
with respect to two well-defined Abelian subalgebras of 
{\it multi-configurational} creation operators $\{C_I^{+}\}$ 
and their Hermitian-adjoint destruction counterparts $\{ C_I^{-} \equiv 
(C_I^{+})^\dagger \}$. The determination of the correlation coefficients 
$\{ {\cal S}_I, \tilde{{\cal S}}_I \}$ is achieved by 
by requiring the ground-state energy expectation 
functional $\bar{H} ( \{ {\cal S}_I, \tilde{{\cal S}}_I\} )$, 
to be stationary with respect to variations in 
each of the (independent) variables of the full set. 
We thereby easily derive the following coupled set of 
equations, 
%%%%%%%%%%%%%%%%% 
\begin{eqnarray} 
\delta{\bar{H}} / \delta{\tilde{{\cal S}}_I} =0 & \Rightarrow &   
\langle\Phi|C_I^{-} {\rm e}^{-S} H {\rm e}^S|\Phi\rangle = 0 ,  \;\; I \neq 0 
\;\; ; \label{ccm_eq7} \\ 
\delta{\bar{H}} / \delta{{\cal S}_I} =0 & \Rightarrow & 
\langle\Phi|\tilde{S} {\rm e}^{-S} [H,C_I^{+}] {\rm e}^S|\Phi\rangle 
= 0 , \;\; I \neq 0 \; . \label{ccm_eq8}
\end{eqnarray}  
%%%%%%%%%%%%%%%% 
Equation (\ref{ccm_eq7}) also shows that the ground-state energy at the stationary 
point has the form 
%%%%%%%%%%%%%%%%
\begin{equation} 
E_g = E_g ( \{{\cal S}_I\} ) = \langle\Phi| {\rm e}^{-S} H {\rm e}^S|\Phi\rangle
\;\; . 
\label{ccm_eq9}
\end{equation}  
%%%%%%%%%%%%%%%% 

It is important to realize that this (bi-)variational 
formulation does {\it not} lead to an upper bound for 
$E_g$ when the summations for $S$ and $\tilde{S}$ in 
Eq. (\ref{ccm_eq2}) are truncated, due to the lack of 
exact Hermiticity when such approximations are made.
However, one can prove  that the important 
Hellmann-Feynman theorem {\it is} preserved 
in all such approximations. 
We utilize various approximation schemes within $S$ and $\tilde{S}$. 
The three most commonly  employed schemes previously utilized 
have been: (1) the SUB$n$ scheme, in which all correlations 
involving only $n$ or fewer spins are retained, but no 
further restriction is made concerning their spatial 
separation on the lattice; (2) the SUB$n$-$m$  
sub-approximation, in which all SUB$n$ correlations 
spanning a range of no more than $m$ adjacent lattice 
sites are retained; and (3) the localized LSUB$m$ 
scheme, in which all multi-spin correlations over 
all distinct locales on the lattice defined by $m$ 
or fewer contiguous sites are retained. 

We now solve the CCM equations in parallel by rearranging 
Eqs. (\ref{ccm_eq7}) and (\ref{ccm_eq8}) for the ket and 
bra states, respectively, where
%%%%%%%%%%%%%%%%% 
\begin{eqnarray} 
{\cal S}_I & = & f_I( ~ {\cal S}_1, ~ \cdots, ~{\cal S}_{I-1}, 
~{\cal S}_{I+1}, ~\cdots,  ~{\cal S}_{N_f} )  \;\; ; \label{ccm_eq10} \\ 
\tilde{{\cal S}}_I & = & \tilde f_I ( ~ {\cal S}_1, ~ \cdots,  ~{\cal S}_{N_f} ;  \nonumber \\
& & ~~~~~~~~ \tilde {\cal S}_1, ~ \cdots, ~\tilde {\cal S}_{I-1}, 
~\tilde {\cal S}_{I+1}, ~\cdots,  ~\tilde {\cal S}_{N_f} ) 
\;\; ; \label{ccm_eq11} 
\end{eqnarray} 
and where $N_f$ is the number of fundamental configurations.
There are therefore $N_f$ equations for both $f_I$ 
and $\tilde f_I$, which refer to the ket and bra states, 
respectively. Each equation contains a finite-number 
of terms and that, indeed, this is always the case for the CCM 
for Hamiltonians that contain a finite number of creation and 
destruction operators. Our approach is to iterate directly the 
set of equations in (\ref{ccm_eq10}) to convergence first in 
order to obtain the ket-state correlation coefficients. We 
then iterate directly the set of equations in (\ref{ccm_eq11}) to 
convergence in order to obtain values for the bra-state 
correlation coefficients. 
\begin{figure}
\epsfxsize=7cm
\epsffile{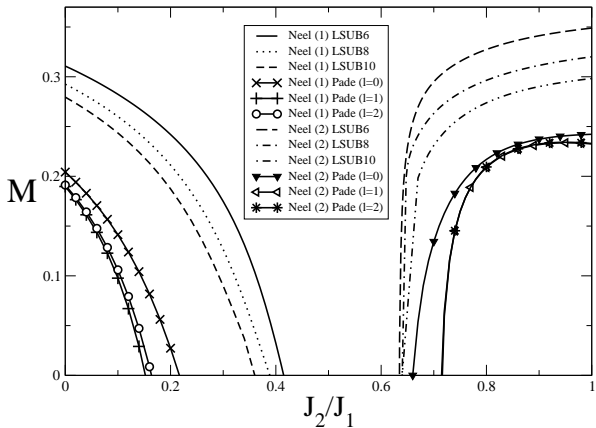}
\vspace{-3.75cm}
\caption{CCM results for the sublattice magnetization of the 
$J_1$--$J_2$ model using the LSUB$m$ approximation 
with $m=\{6,8,10\}$. Extrapolations using Pad\'e approximants 
are also shown.}
\label{fig4}
\end{figure}

\begin{table}[t]
\caption{CCM LSUB$m$ results for the CAVO antiferromagnet. The 
numbers of fundamental configurations nearest-neighbor and 
next-nearest-neighbor N\'eel model states are given by
$N_f^{n.n.}$ and $N_{f}^{n.n.n.}$, respectively. Results for 
the points at which the sublattice magnetization goes to 
zero for the nearest-neighbor and next-nearest-neighbor 
N\'eel model states are indicated by $J_2/J_1|_1$ and 
$J_2/J_1|_2$, respectively. (Errors for the critical 
points are in the last decimal place shown.) 
Ground-state energies and 
sublattice magnetizations for nearest-neighbor bonds 
only, namely at $J_1=1$ and $J_2=0$, are extrapolated
using Pad\'e approximants for $m=\{4,6,8,10\}$, where 
$l$ is the order of the polynomial in the denominator.
By way of comparison, results of exact diagonalizations 
(ED) for $E_g/N$ and $M$ of Ref. \cite{richter} and 
values for $J_2/J_1|_1$ and $J_2/J_1|_2$ of Ref. \cite{albrecht} 
are also given.}
\begin{tabular}{|l|c|c|c|c|c|c|}  \hline\hline
  Method  &$N_f^{n.n.}$  &$E_g/N$  &$M$ 
          &$N_{f}^{n.n.n.}$  &$J_2/J_1|_1$ &$J_2/J_1|_2$       \\ \hline\hline
LSUB2	  &2       &$-$0.52859   &0.393   &2     &0.73   &0.29   \\ \hline
LSUB4	  &14      &$-$0.54595   &0.339   &18    &0.48   &0.60   \\ \hline
LSUB6	  &130     &$-$0.54984   &0.311   &208   &0.42   &0.63   \\ \hline
LSUB8	  &1589    &$-$0.55129   &0.293   &2715  &0.39   &0.64   \\ \hline
LSUB10	  &22395   &$-$0.55200   &0.280   &39434 &0.36   &0.65   \\ \hline\hline       
Pad\'e $l=0$  &--  &$-$0.55332	 &0.204     &--    &0.22   &0.66   \\ \hline
Pad\'e $l=1$  &--  &$-$0.55338   &0.190     &--    &0.16   &0.72   \\ \hline
Pad\'e $l=2$  &--  &$-$0.55340   &0.191     &--    &0.18   &0.71   \\ \hline\hline       
ED  \cite{albrecht,richter}      &--      &$-$0.5533    &0.2303 
                                 &--    &0.2 &0.7     \\ 
\hline\hline
\end{tabular} 
\label{tab1}
\end{table}

Our parallelization technique is 
now to split the problem of determining and solving the 
$N_f$ equations for both the bra and ket states of Eqs. 
(\ref{ccm_eq10}) and (\ref{ccm_eq11}) between each 
processor in the parallel cluster. Thus, each equation 
for $I=\{1,\cdots,N_f\}$ is ever dealt with by one node 
in the cluster only and the results of each node are 
collected together at each iteration. (Again, we remember 
that we iterate the ket-state equations first and then the 
bra-state equations.) Thus, the set of equations for the 
ket and bra states are solved independently and in parallel.
Eight Compaq ``Wildfire'' processors were used here, 
each with 2 GIGs of RAM and CPU speeds of 731 MHz. The 
run for LSUB10 with the nearest-neighbor model state
with 22395 fundamental configurations took 11 days for
18 distinct values of $J_2/J_1$.

We now wish only to illustrate this approach by considering 
the the CAVO system of Eq. (\ref{eq1}). The ground-state 
energies of the pure nearest-neighbor antiferromagnet 
(at $J_1=1$ and $J_2=0.0$)  are shown in Table \ref{tab1}. 
Fig. \ref{fig2} shows LSUB$m$ ground-state energy plotted 
against $m^{-2}$ for $m=\{4,6,8,10\}$.
``Linear extrapolations'' of data points have previously been
used to great effect for isotropic nearest-neighbor
antiferromagnets \cite{ccm1,ccm2,ccm3,ccm4,ccm5}. In this case
of the CAVO model studies here, we obtain a value of 
$E_g/N=-0.55310(5)$. We also carry out extrapolations 
using Pad\'e approximants \cite{ccm3} for $E_g/N$ plotted
against $m^{-2}$ for $m=\{4,6,8,10\}$. $l$ 
is the order of the polynomial in the denominator and that
we perform an exact fit of the curve to the data. 
Hence, in contrast to the simple linear fit, there is no
``curve-fitting'' error. Extrapolations using Pad\'e 
approximants for the ground-state energy at $J_1=1$ and 
$J_2=0.0$ are found to be in excellent agreement with a 
variational QMC result \cite{miyazaki} of $E_g/N=-0.5501$ 
and an ED result \cite{richter} of $E_g/N=-0.5533$. We conclude 
that our results are converged to at least 3 decimal places, 
and we believe that these results constitute one of the 
most accurate estimations of the ground-state energy for 
this system ever found. 

The number of fundamental configurations for the SUB10-10 approximations 
were 22340 and 39434 for the nearest-neighbor and 
next-nearest-neighbor model states, respectively, and 
this calculation for the next-nearest-neighbor
model state is the largest CCM calculation as yet performed 
for quantum antiferromagnetic spin systems by an order of 
magnitude compared to the previous highest \cite{ccm4}. 
The results for the ground-state energies for 
both model states are shown in Fig. \ref{fig3} with respect 
to $J_2/J_1$. We see that these results converge quickly 
with truncation index $m$ for the LSUB$m$ approximation 
over a wide range of $J_2/J_1$.  The wide area
without results is an indication ground-state ordering (e.g.,
dimer or plaquette) not accessible with our N\'eel model states.

LSUB$m$ results for the sublattice magnetization 
at $J_1=1$ and $J_2=0$ are again shown in Table \ref{tab1}.
In contrast to the case for the ground-state energy, 
extrapolations for the sublattice magnetization do not 
seem to follow a similar linear extrapolation ``rule''
as shown in Fig. \ref{fig4}. However, as in Refs.
\cite{ccm1,ccm2,ccm3,ccm4,ccm5} we may plot the sublattice
magnetization against $m^{-1}$ and use Pad\'e approximants
in order to carry out the extrapolations. This gives
values for $M$ in the range 0.19 to 0.20, which is 
again in good agreement with the best of other
approximate methods, such as results from
QMC \cite{troyer} and ED \cite{richter} of $M=0.178(8)$
and $M=0.2303$, respectively. We therefore use Pad\'e
approximants for increasing values of $J_2/J_1>0$
as a reliable indicator of the position of the phase 
transition points is given by the position at 
which the sublattice magnetization goes to zero
as a function of $J_2/J_1$. Fig. \ref{fig4} plots our 
high-order CCM results for the amount of sublattice 
magnetization. We may see from Table \ref{tab1}
that our results for the positions of the phase
transition points are in good agreement with
results of exact diagonalizations \cite{albrecht} that predict 
that plaquette and dimer ordering exists in the
region $0.2 < J_2/J_1 < 0.7$.

We have shown in this article that it is possible 
to simulate the properties of a CAVO material at
zero temperature using parallelized high-order CCM 
techniques. The construction 
of CCM codes to such high orders of approximation is 
a non-trivial task. Furthermore, the parallelization of 
the CCM code adds an additional level of complexity to it. 
The 39434 fundamental configurations for the 
next-nearest-neighbor N\'eel model state is a 
massive leap forward in the practical application 
of the CCM, which is impossible without an efficient 
parallelization. However, the numbers of 
fundamental configurations of order 10$^6$ for quantum 
antiferromagnets are clearly within our reach 
via parallel processing for even larger clusters 
of processors than the relatively modest number of 
(eight) processors used here. We have therefore 
``proven the principle'' that the CCM 
can be effectively parallelized, and we believe
that we have also shown here that this, in turn,
leads to great increases in accuracy of observed
expectation values.

The extension of these calculations to treat 
quantum systems that demonstrate ``novel ordering,''
such as with plaquette or dimer solid ground states,
is, in principle, straightforward. The basic building
blocks for the CCM using dimer models states are 
already in place, and low-order calculations
have already been attempted successfully for a 
one-dimensional frustrated spin system \cite{xian}. 
However, procedurally, the spatial dimension
of the lattice makes no difference to the coding of
the problem, although the computational expense  
grows with it and so efficient parallelized approaches,
such as that shown here, would be very useful in this
case. Extensions of the method to consider excitation
energies have been very successful \cite{ccm3}, and 
it is again a straightforward matter to obtain 
excitation spectra using ``localized'' high-order
CCM approximations. The CCM may be applied to consider
systems even in the presence of strong frustration
and for lattices of arbitrary spatial dimensionality, 
and the CCM now constitutes a very powerful and a useful 
tool in order to understand the basic properties of 
quantum spin systems.

%%%%%%%%%
%bibliography%%
%%%%%%%%%


\begin{thebibliography}{200}

% v CCM theory v

\bibitem{ccm_theory1} R.F. Bishop, {\it Theor. Chim. Acta} {\bf 80}, 95 (1991). 

% ^ CCM theory ^

% v CCM applied to spin systems v 

\bibitem{ccm1} R.F. Bishop, J.B. Parkinson, and Yang Xian, 
{\it Phys. Rev. B} {\bf 46}, 880 (1992). 

\bibitem{ccm2} C. Zeng, D.J.J. Farnell, and R.F. Bishop, 
{\it J. Stat. Phys.}, {\bf 90}, 327 (1998).

\bibitem{ccm3} R. F. Bishop, D. J. J. Farnell, S.E. Kr\"uger, J. B. 
 Parkinson, J. Richter, and C. Zeng, {\it J. Phys.: Condens. Matter}
 {\bf 12}, 7601 (2000).


\bibitem{ccm4} S.E. Kr\"uger, J. Richter, J. Schulenburg, 
 D.J.J. Farnell, and R.F. Bishop, {\it Phys. Rev. B} {\bf 61},
 14607 (2000).

\bibitem{ccm5} D. J. J. Farnell, K. A. Gernoth, and R. F. Bishop, 
{\it J. Stat. Phys.} {\bf 108}, 401 (2002).

\bibitem{ccm6} D. J. J. Farnell and R. F. Bishop, in 
 ``Quantum Magnetism,'' {\it Lecture Notes In Physics} {\bf 645}. 
 (Springer Verlag, Berlin, Heidelberg, 2004), p. 307-348. 

% yang xian ccm for dimer model states

\bibitem{xian} Y. Xian, {\it J. Phys.: Condens. Matt.} {\bf 6}, 5965
 (1994).

% yang xian ccm for dimer model states


% ^ CCM applied to spin systems ^

% v previous results for cavo antiferromagnet v

% LSWT treatment of j1-j2 cavo model using neel, dimer, and plaquette. Shows pt boundaries of 0.25 and 0.8. 

\bibitem{starykh} O.A. Starykh, M.N. Zhitomirsky, D.I. Khomskii, R.R.P. Singh, and K. Ueda, {\it  Phys. Rev. Lett.} {\bf 77}, 2558 (1996).

% qmc for cavo n.n. only. m=0.178(8)

\bibitem{troyer} M. Troyer, H. Kontani, and K. Ueda, {\it Phys. Rev. Lett.} {\bf 76}, 3822 (1996).

% exact diag up to 32 sites for j1-j2 model. gives PTs at 0.2 and 0.7.

\bibitem{albrecht} M. Albrecht, F. Mila, and D. Poilblanc, {\it Phys. Rev. B} {\bf 54}, 15856 (1996).

%DMRG 2D results indicating gap opens up at 0.05.

%\bibitem{white} S.R. White, {\it Phys. Rev. Lett.} {\bf 77}, 3633 (1996).

% plaquette series expansions for j-j'-j2 model
%some results for j=j' indicating p.t. at j2/j1=0.1 to 0.2.

\bibitem{weihong1} Z. Weihong, M.P. Gelfand, R.P. Singh, {\it Phys. Rev. B} {\bf 55}, 11377 (1997).

%excitation gaps using series expansions.

\bibitem{weihong2} Z. Weihong, J. Oitmaa, and C.J. Hamer, {\it Phys. Rev. B} {\bf 58}, 14147 (1998). 

%variational approach. some bizarre results really. still plaquette and dimer.

\bibitem{ueda} K. Ueda, H. Kontami, M. Sigrist, and P.A. Lee, {\it Phys. Rev. Lett.} {\bf 76}, 1932 (1996).

% RVB variational: energy= -0.5501

\bibitem{miyazaki} T. Miyazaki and D. Yoshioka, {\it cond-mat/9602133}.

% ED by johannes e=-0.5335 & M=0.2303

\bibitem{richter} J. Richter, J. Schulenburg, and A. Honecker, in 
 ``Quantum Magnetism,'' {\it Lecture Notes In Physics} {\bf 645}. 
 (Springer Verlag, Berlin, Heidelberg, 2004), p. 106. 

\end{thebibliography}
\end{document}